\documentclass[aps,prl,reprint,superscriptaddress]{revtex4-1}

\usepackage{booktabs}
\usepackage{braket}
\usepackage{graphicx}%
\usepackage{dcolumn}%
\usepackage{bm}%
\usepackage[per-mode=symbol]{siunitx}
\usepackage{units}
\usepackage{color}
\usepackage{amsmath}
\usepackage{amssymb}

\DeclareMathOperator{\Tr}{tr}
\newcommand\beq{\begin{equation}}
\newcommand\eeq{\end{equation}}
\newcommand\bfig{\begin{figure}}
\newcommand\efig{\end{figure}}

\begin{document}

\title{Obtaining tight bounds on higher-order interferences with a 5-path interferometer}

\author{Thomas Kauten}
\email{thomas.kauten@uibk.ac.at} 
\author{Robert Keil}
\author{Thomas Kaufmann}
\author{Benedikt Pressl}
\affiliation{Institut f\"ur Experimentalphysik, Universit\"at Innsbruck, Technikerstra\ss{}e 25, 6020 Innsbruck, Austria}
\author{\ifmmode \check{C}\else \v{C}\fi{}aslav Brukner}
\affiliation{Faculty of Physics, University of Vienna, Boltzmanngasse 5, A-1090 Vienna, Austria}
\affiliation{Institute for Quantum Optics and Quantum Information (IQOQI), Boltzmanngasse 3, A-1090 Vienna, Austria}
\author{Gregor Weihs}
\affiliation{Institut f\"ur Experimentalphysik, Universit\"at Innsbruck, Technikerstra\ss{}e 25, 6020 Innsbruck, Austria}

\begin{abstract}
Within the established theoretical framework of quantum mechanics, interference always occurs between pairs of trajectories. Higher order interferences with multiple constituents are, however, excluded by Born's rule and can only exist in generalized probabilistic theories. Thus, high-precision experiments searching for such higher order interferences are a powerful method to distinguish between quantum mechanics and more general theories. Here, we perform such a test in optical multi-path interferometers. Our results rule out the existence of higher order interference terms to an extent which is more than four orders of magnitude smaller than the expected pairwise interference, refining previous bounds by two orders of magnitude. This establishes the hitherto tightest constraints on generalized interference theories.
\end{abstract}

\maketitle

Since arising almost a century ago, quantum mechanics has long become an established paradigm for the description of nature on a submicroscopic scale. It is at the basis of an enormous variety of present and potential future applications, such as quantum communication \cite{duan2001,Gisin:QuantumCommunicationReview}, quantum computation \cite{NielsenChuang:QCBook,knill2001,Ladd2010} and protocols like entanglement swapping \cite{pan98} or teleportation \cite{boschi98}. However, all these applications rely ultimately on interference and entanglement, which can be alternatively explained by theories sharing only some fundamental features with quantum mechanics, such as the superposition principle or probabilistic predictions for outcomes, and yet differing from it in other aspects. In order to distinguish between quantum theory and such alternatives one needs to design dedicated experiments. The situation may be compared with the time before the first Bell test experiments had been performed. Until then one could explain all quantum mechanical phenomena with a local hidden variable theory. It was required to first state Bell's theorem \cite{bell64} and then to perform dedicated experiments with space-like separated laboratories to exclude the alternative. Only last year all experimental loopholes were finally closed \cite{bell2, bell3, bell4}. Another example is the experiment that distinguishes between quaternion and complex (standard) quantum theory \cite{kaiser84, procopio16}. Here we focus on an experimental test capable of discerning between quantum mechanics and its generalizations exhibiting higher order interference.

The probabilistic nature of quantum theory is stated by Born's rule \cite{born1926}, i.e. that the probability density $P(\mathbf{r},t)$ for an observation of a quantum object at a certain time $t$ and a certain position $\mathbf{r}$ is given by the absolute square of its wavefunction $\Psi(\mathbf{r},t)$:
\begin{equation} P(\mathbf{r},t)=\Psi^*(\mathbf{r},t)\Psi(\mathbf{r},t)=\left|\Psi(\mathbf{r},t)\right|^2.
\label{BornsRule}
\end{equation}
As a consequence of Born's rule and quantum superposition, interference can take place even for single particles \cite{Joensson:DoubleSlitElectrons}. For concreteness, consider an interferometer with multiple non-overlapping paths $k=A,B,C,\ldots$ which superpose  in some output port to the final wavefunction $\Psi=\sum_k\Psi_k$. Eq.~(\ref{BornsRule}) implies:
\begin{equation} P(\mathbf{r},t)=\sum_k\left|\Psi_k(\mathbf{r},t)\right|^2+\sum_{k<l}I_{kl}(\mathbf{r},t), \label{eq:Prt}\end{equation}
with pairwise (first-order) interference terms $I_{kl}\equiv\Psi_k\Psi_l^{\ast}+\mathrm{c.c.}$, depending on the relative phase between the two paths $k$ and $l$. Thus, one obtains interference terms that always originate from pairings of paths, but no higher order interferences involving more than two paths at once.

In this vein, one can use the presence or absence of higher-order interferences as an experimental probe of the current framework of quantum mechanics. First developed by Sorkin in the context of a measure theory on spacetime \cite{sorkin1994}, one can define a hierarchy of interference terms. In a 3-path interferometer with individually blockable paths $A,B,C$,  where $P_{ABC}$ is the probability to find a particle in the output port of the interferometer if all paths are open, $P_{AB}$ for only paths $A$ and $B$ being open, etc. The so-called second-order interference term
\begin{equation} I_{ABC}\equiv P_{ABC}-P_{AB}-P_{AC}-P_{BC}+P_A+P_B+P_C \label{eq:iabc}\end{equation}
should be zero, independent of the individual phases and powers in each interferometer arm, due to Eq.~\ref{eq:Prt}. Conversely, a significant deviation from $I_{ABC}=0$ would indicate the existence of higher-order interferences and contradict conventional quantum theory. Note that the definition~(\ref{eq:iabc}) accounts for deviations from the standard theory in a model-independent way.

In any experiment with discrete particles, the probability $P$ will be proportional to the detected particle flux $p$. Therefore a directly measurable quantity
\begin{equation} \epsilon_3\equiv p_{ABC}-p_{AB}-p_{AC}-p_{BC}+p_A+p_B+p_C-p_0\end{equation}
can be defined \cite{sinha08}. In this expression, for example $p_{AB}$ is the detected particle flux at the output when only paths $A$ and $B$ are open. The background term $p_0$ gives the measured signal when all paths are blocked, accounting for detector dark current/dark counts. For better comparison of the results with the expected behavior, one can introduce the normalized quantity $\kappa_3\equiv\epsilon_3/\delta_3$ measuring the ratio of hypothetical second-order interference to the sum of the expected first-order interference,  $\delta_3\equiv|I_{AB}|+|I_{AC}|+|I_{BC}|$. For interferometers with more than three paths, the higher (third, fourth,...)-order interference terms ($\kappa_4$, $\kappa_5$,...), which of course are also zero in standard quantum theory, can be defined accordingly. 

Different experiments have been realized previously to obtain an upper bound on the modulus of the second-order interference term. These experiments were implemented in optics \cite{sinha10,hickmann11,sollner12} as well as via nuclear magnetic resonance (NMR) in molecules \cite{park12}, delivering results which were all in accordance with the expectation $\kappa_3=0$. The NMR experiment provided the hitherto tightest constraint with $\kappa_3 = 0.001 \pm 0.003$.

As is the case for any such null-test experiment, the tightness of the bound and, thereby, the strength of any conclusions to be drawn about the foundations of the theory depend on the measurement uncertainties. In previous optical 3-path interferometers, the precision was mostly limited by the phase stability of the interferometer, while the accuracy suffered from detector nonlinearities \cite{sollner12}. In this work, we present a greatly improved multi-path experiment, namely a stabilized 5-path interferometer with single photons, with which we are not only able to tighten the bound on second-order interference by two orders of magnitude, but also measure third and fourth-order interference terms. The 5-path interferometer has the additional advantage of permitting the acquisition of more statistics for the second- and third-order interference term since it consists of ten 3-path interferometers and five 4-path interferometers. The systematic error of detector nonlinearities is taken into account by separate detector calibration and full quantum state tomography of the produced 5-dimensional qudit state.

\begin{figure}
\centering
\includegraphics[width=\columnwidth]{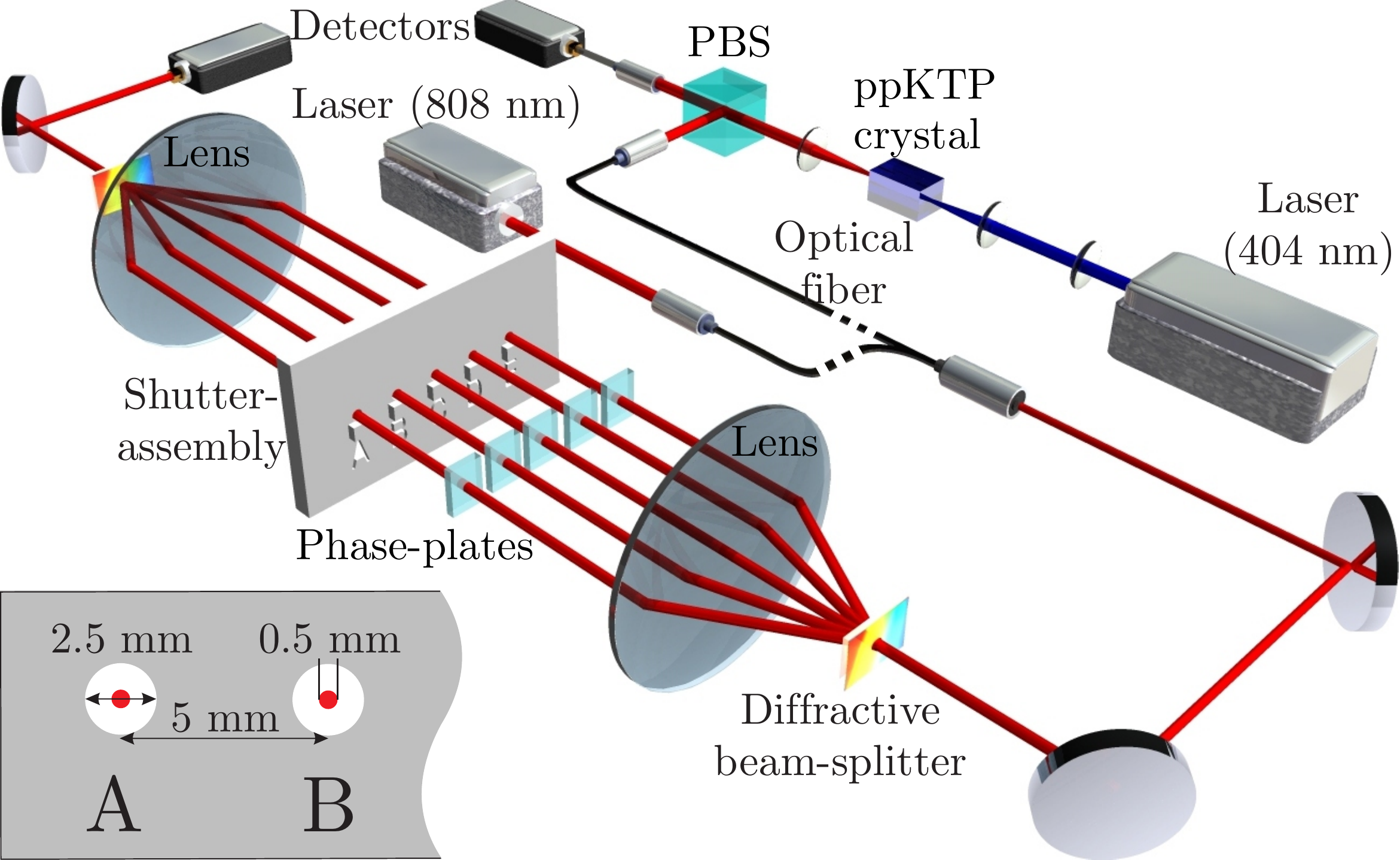}
\caption{Experimental setup. The light source is either given by a power stabilized laser or single photons (both at \SI{808}{\nano\meter}) produced via SPDC in a ppKTP crystal pumped by a blue laser. The interferometer consists of two diffracitve beamsplitters and two lenses. Shutters and phase plates in each of the paths allow independent manipulation. The inset shows the dimensions and separations of the shutters and the $1/e^2$ intensity diameters of the beams.}
\label{fig:setup}
\end{figure}

A schematic drawing of our setup can be seen in Fig.~\ref{fig:setup}. We employed light sources in three different regimes: classical (C), semi-classical (SC) and quantum (Q). In the classical regime we used a continuous-wave single frequency laser at \SI{808}{\nano\meter} power-stabilized to \SI{1}{\milli\watt} with relative fluctuations smaller than 0.1\% over the complete measurement time of several days by using a liquid crystal noise eater (Thorlabs LCC3112).  We used a single photon source to perform measurements in the other regimes. Photon pairs at 808nm are produced via type-II spontaneous parametric down conversion (SPDC) in a \SI{10}{\milli\meter} long periodically poled potassium titanyl phosphate (ppKTP) crystal which is pumped by a blue laser (\SI{404}{\nano\meter}). The orthogonally polarized photons are separated on a polarizing beam splitter (PBS). We collect $6\times 10^5$ single photons per second in each of the outputs in single mode fibers and we get $10^5$ pairs per second at \SI{4}{\milli\watt} pump power. One of the photons serves as a heralding photon, whereas the other is sent through our multi-path interferometer. Therefore we have two possibilities to conduct the measurement with single photons: either free running, where all photons transmitted through the interferometer are counted (yielding a thermal photon number distribution) in the semi-classical regime or conditioned, where only photons are counted if there is a heralding photon (producing a sub-Poissonian distribution) \cite{hsps1} in the quantum regime. All light sources were linearly polarized.

The interferometer is a Mach-Zehnder 5-path interferometer consisting of a diffractive beam splitter (a diffractive optical element -- Holoeye DE 263 -- modulating the incident light via a micro-relief surface) which creates five almost equally powerful beams, collimated by a lens ($f=\SI{150}{\milli\meter}$). A shutter assembly serves to block or unblock each of the five beams individually, phase plates (glass plates with a thickness of \SI{0.15}{\milli\meter} and anti-reflection coated for \SI{808}{\nano\meter}) mounted on motorized rotation stages in all of the five beams allow us to set the phase of each path independently. The absolute angular repeatability is \SI{0.005}{\degree}, which corresponds to $\pi/1000$ in phase. An identical pair of lens and diffractive element sends the resulting beam onto a detector. The interferometer is designed in $4f$-configuration and the individual beams are separated by \SI{5}{\milli\meter}, therefore the overall dimensions are $(60\times 2)$~cm$^2$. For detecting single photons we used SPCM-AQRH-12-FC single photon counting modules from Perkin Elmer followed by a quTAU time-to-digital converter from qutools GmbH. This system has a deadtime of $(33.85\pm 0.31)~$\SI{}{\nano\second} and $(150\pm 18)$ dark counts per second. The laser radiation is detected by a Physimetron A139-001 photoreceiver based on a Si-photodiode (Hamamatsu S2386-18K) and a \SI{1}{\mega\volt\per\ampere} transimpedance amplifier, read out by an Agilent 34410A multimeter. This detection system has a low maximum nonlinearity of less than $35$~ppm \cite{kauten14}. One measurement set consists of the $2^5=32$ different possible open/close combinations of the five paths. These combinations were measured in random order to reduce the influence of any memory effects of the detector and of drifts of the source. To obtain data with comprehensive statistics we recorded several thousand measurement sets within a total measurement time of several days. The whole interferometer is shielded against air motion and stray light as well as passively and actively temperature stabilized with a PI controller (Wavelength Electronics HTC1500) and heating mats to a root-mean-square fluctuation $<\SI{0.02}{K \per 24 \hour}$ (The temperature was monitored with a PT1000 resistance thermometer). Additionally, the phases are actively stabilized by optimizing the phase-plate position after 100 measurement cycles towards maximally constructive interference of all two-path combinations. This point in phase space were chosen for convenience of alignment and because small phase changes lead only in second order to deviations in output power. This results in good phase stability over the whole measurement time. By comparing the fluctuations and drifts of the single-path power with the multi-path powers (see Fig.~4 and Fig.~5 of the Supplemental Material \cite{SuppS5}), which have the same order of magnitude, we found that phase uncertainty plays a minor role compared to power noise from the light source.

\begin{figure}
\centering
\includegraphics[width=\columnwidth]{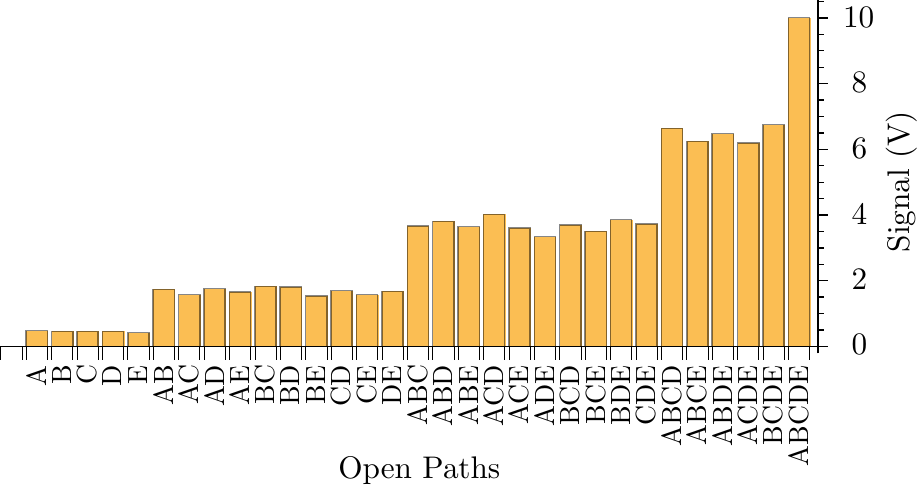}
\caption{Mean values of the measured powers for the different path-combinations with classical light. Extreme outliers have already been removed.}
\label{fig:intensity}
\end{figure}

The resulting average powers of the different path combinations can be seen in Fig. \ref{fig:intensity} for the measurement with the power stabilized laser \cite{SuppS5}. We filtered the data for extreme outliers (resulting from shutter failure) according to Grubbs' test for outliers (with a significance level of $99\%$) \cite{grubbs50,grubbs69}. After that the largest relative standard deviation of the various classical signals is $0.3\%$ for 5618 measurement sets recorded within \SI{68}{\hour}. For the semi-classical (quantum) single photon measurement with 1912 measurement sets the largest standard deviation was measured to be $3.6\%$ ($15.5\%$) over a measurement time of \SI{88}{\hour}. These higher values result mainly from shot noise and from the fact that the power of the blue pump laser was not stabilized.

\begin{figure}
\centering
\includegraphics[width=\columnwidth]{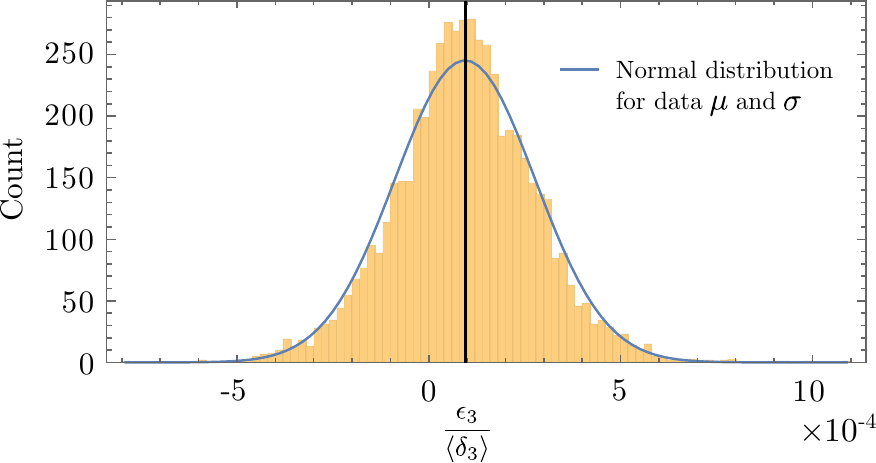}
\caption{Histogram of $\epsilon_3/\langle\delta_3\rangle$ values in the classical regime for the 3-path subset $\left\{A,B,C\right\}$. The blue line is a Gaussian fit to the distribution of the data and the mean value is indicated by the vertical black line, whose width corresponds to the standard error of the mean (one standard deviation).}
\label{fig:hiso}
\end{figure}

Due to the anti-correlation between the numerator $\epsilon$ and the denominator $\delta$ in the definition of $\kappa$, a bias towards positive values can arise from random fluctuations in the data when calculating $\kappa$ for every shutter cycle and averaging over the data sets. However, calculating the averages of numerator and denominator in the definition of $\kappa$ separately,
\begin{equation} \langle \kappa_j\rangle \equiv\frac{\langle\epsilon_j\rangle}{\langle\delta_j\rangle}, \qquad\:j=3,4,5 \end{equation}
eliminates their correlations and yields an unbiased estimator of the higher-order interference terms. Indeed, one can show that error sources, which typically occur in interference experiments, such as power fluctuations of the photon source, countrate fluctuations of the detectors (Poissonian photon counting uncertainties), detector/electronic noise, coherent phase fluctuations as well as incoherence, have no systematic effect on the measurement outcome \cite{SuppS1}.

For each of the measured 32-tuples we calculate $\epsilon_{3,4,5}$ and $\delta_{3,4,5}$. A histogram plot of the measured ensemble of $\epsilon_3/\langle\delta_3 \rangle$ in the classical regime is shown in Fig. \ref{fig:hiso} \cite{SuppS2}. After averaging across all possible path combinations one obtains the mean values and associated uncertainties presented in table \ref{tab:exp}.

\sisetup{
table-number-alignment=center,
separate-uncertainty=true,
table-figures-integer = 3,
table-figures-decimal = 1}

\begin{table}[!ht]
\centering
\begin{tabular}{l||
                S[separate-uncertainty,table-figures-uncertainty=1]|
                S[separate-uncertainty,table-figures-uncertainty=1]|
                S[separate-uncertainty,table-figures-uncertainty=1]}
\toprule
             
{}    & {$\langle\kappa_3\rangle$} & {$\langle\kappa_4\rangle$} & {$\langle\kappa_5\rangle$}  \\
\midrule \hline
{classical $(\times 10^{-5})$}&9.7 \pm 0.1&2.7 \pm 0.2&0.3 \pm 0.3\\ 
{semi-classical ($\times 10^{-4}$)}&-9.9 \pm 1.8&-5.1 \pm 2.1&-3.8 \pm 3.9\\  
{quantum $(\times 10^{-3})$}&-1.1 \pm 1.6&0.3 \pm 1.8&-2.6 \pm 2.9\\  
\bottomrule
\end{tabular}
\caption{Mean values of the measured higher-order interferences and their standard errors in the classical, semi-classical and quantum regimes.}
\label{tab:exp}
\end{table}

\begin{figure}
\centering
\includegraphics[width=\columnwidth]{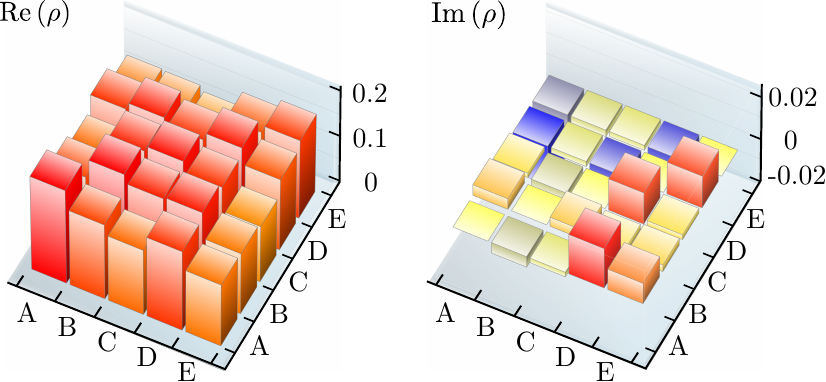}
\caption{The density matrix $\rho$ of the 5-dimensional qudit state in our interferometer, with $\Tr\rho^2=0.74$.}
\label{fig:rho}
\end{figure}

It was recently shown that near-field effects in slit-based measurements, where the relevant dimensions are just one or two orders of magnitude larger than the wavelength $\lambda$, can lead to an apparent higher-order interference and, therefore, bias the experiment \cite{DeRaedt2012,sinha14}. However, in our interferometer these effects are of negligible influence, due to the macroscopic separation of the paths and beam width, exceeding $\lambda$ by 3 to 4 orders of magnitude (see inset in Fig. \ref{fig:setup}). Instead, the main systematic uncertainty in our experimental configuration arises from the nonlinearity of the detectors. Real detectors usually have a nonlinear response function, which means that the recorded value (voltage, photon counts,...) is not linear in the incident power or photon flux, but biased differently for different optical powers. This biases the value of $\kappa$, as we measure light powers varying over more than one order of magnitude. Here, the bias arises mainly due to nonlinearities in the electronics of our photoreceiver and due to deadtime in the single photon detector. To take this error into account it is useful to fully characterize our 5-path interferometer, which can be described as a 5-dimensional qudit state. Therefore, we additionally performed complete quantum state tomography \cite{thew02,Altepeter05}. The density matrix $\rho$ was numerically reconstructed from single- and two-path measurements with defined phases via direct reconstruction. The phases are calibrated via scanning the classical two-path laser interference. We used the direct reconstruction instead of a maximum likelihood estimation to avoid systematic deviations in the state reconstruction, which have recently been shown to arise due to the constraint of physicality in maximum likelihood estimates \cite{schwemmer15}. The real and imaginary parts of the resulting density matrix are shown in Fig. \ref{fig:rho}. We calculated $\Tr\rho^2=0.74$; the deviation from 1 (a pure state with no which-path information) can be attributed to an imperfect overlap of the five beams at the second beamsplitter. While this degree of coherence in the interferometer must be determined for an accurate prediction of the influence of the nonlinearities, its actual value has no systematic impact on the Sorkin experiment. The effect of the nonlinearity on the reconstruction is negligible for two reasons: the ratio of the photon fluxes for the different measurement settings is much smaller than in the measurements contributing to the evaluation of $\kappa$. More importantly, deviations in the density matrix do not produce a systematic effect on the expected higher-order interference, as $\kappa=0$ holds for all states in quantum theory.

From the density matrix it is possible to calculate the expected powers for the different settings of the shutters in the Sorkin experiment. We found good accordance with our measurement data \cite{SuppS3}, suggesting that the tomography produces an accurate description of the interferometer. The nonlinearities of both detectors have been characterized in separate experiments \cite{kauten14}. Applying them to the powers/count rates predicted from the density matrix yields small corrections of these powers (relative change $<0.03\%$ for the laser powers and $<0.5\%$ for the unheralded single photon rates).

One can then calculate the apparent higher-order interferences $\kappa_{\textnormal{th}}$, which would be expected in the Sorkin measurements, from these corrected data. The differences between the experimentally measured higher-order interferences and the expected values due to the nonlinearities $\tilde{\kappa}\equiv\langle\kappa\rangle -\kappa_{\textnormal{th}}$ give corrected higher-order interferences as the final results, which can be found in table \ref{tab:sim}. 

\sisetup{
table-number-alignment=center,
separate-uncertainty=true,
table-figures-integer = 3,
table-figures-decimal = 1}

\begin{table}[!ht]
\centering
\begin{tabular}{l||
                S[separate-uncertainty,table-figures-uncertainty=1]|
                S[separate-uncertainty,table-figures-uncertainty=1]|
                S[separate-uncertainty,table-figures-uncertainty=1]}
\toprule
             
{}    & {$\tilde\kappa_3$} & {$\tilde\kappa_4$} & {$\tilde\kappa_5$}  \\
\midrule \hline
{classical $(\times 10^{-5})$}&0.0 \pm 3.1&4.3 \pm 4.4 &4.2 \pm 5.1\\
{semi-classical ($\times 10^{-4}$)}&1.3 \pm 1.8&-1.6 \pm 2.1&-3.8 \pm 4.0\\  
{quantum $(\times 10^{-3})$}&0.0 \pm 1.6&0.6 \pm 1.8&-2.7 \pm 2.9\\
\bottomrule
\end{tabular}
\caption{$\tilde{\kappa}\equiv\langle\kappa\rangle-\kappa_{\textnormal{th}}$ is the nonlinearity-corrected higher-order interference for all measurement regimes. All these values are within one standard deviation of the expected zero value.}
\label{tab:sim}
\end{table}

\begin{figure}
\centering
\includegraphics[width=\columnwidth]{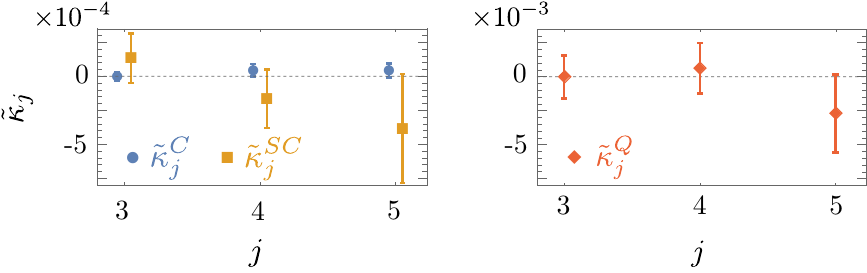}
\caption{Final result. $\tilde\kappa$ gives the difference between the experimentally measured $\langle\kappa\rangle$ and the value expected from detector nonlinearities $\kappa_{\textnormal{th}}$ for the measurements in the three regimes. The order of $\tilde\kappa$ increases along the horizontal axis and the error bars indicate one standard deviation.}
\label{fig:kappacomp}
\end{figure}

Note that in case of the heralded single photon data, we did not calculate an explicit prediction for $\kappa_{\textnormal{th}}$ because the nonlinearity model is quite involved in this case. Instead we used the model to correct the raw experimental data, in order to obtain $\tilde{\kappa}$ \cite{SuppS4}. A final summary of all the different $\tilde\kappa_j$ values is presented in Fig. \ref{fig:kappacomp}. One finds that all these values are within one standard deviation of the expected zero value.

The optical 5-path interferometer presented in this work permitted us to experimentally confine the allowed domain of second order interference to an uncertainty of $3\times 10^{-5}$ in the classical light regime. This is two orders of magnitude tighter than the bounds obtained from the most precise experiments in any system to date. The uncertainties in the semi-classical and quantum regimes of $2\times 10^{-4}$ and $2 \times 10^{-3}$, respectively, are also much lower than what has been reported before \cite{sinha08,sollner12}. This new level of precision has been reached by a range of technical improvements over previous interferometers including power stabilization, phase stabilization and increased throughput as well as a judicious analysis of detector nonlinearities, which are the dominant origin of systematic error. Furthermore, we have performed the first measurement of third- and fourth-order interference terms, with similarly small uncertainties. So far, all our experimental results showed no significant higher-order interferences and are, therefore, in full accordance with the conventional theory. The dominant sources of imprecision in our setup are the uncertainties in determining the detector nonlinearities as well as shot noise in the single photon regime. In order to narrow the bound on higher-order interference further, highly linear detection systems and brighter single photon sources or higher detection efficiency will be required. A narrower experimental bound will aid the development of new theories or constrain free parameters of existing ones. In particular, knowledge of the various higher-order terms should permit discriminating between different models for generalized theories, such as coefficients in nonlinear extensions of Born's rule \cite{Khrennikov:FourthPowerBornRule}, the theory of density cubes \cite{dakic14} and quartic quantum theory \cite{Zyczkowski2008}. For example, the bounds on $\tilde{\kappa}_3$ translate directly to bounds on the magnitude of off-diagonal elements in the theory of density cubes \cite{dakic14}. All these alternative theories contain quantum theory as a subset similarly as quantum theory contains classical theory as a subset. The mechanism by which theories exhibiting higher-order intereferences reduce to standard quantum theory is called hyper-decoherence \cite{Zyczkowski2008, lee2016}. This mechanism would be analogous to the process of decoherence, which induces the quantum-to-classical transition. Our experiment places also a bound on the hyperdecoherence time of the potential extensions of quantum theory with second, third and fourth order interference. Such post-quantum theories are not only interesting from the foundational point of view; they could solve problems intractable even on a quantum computer \cite{lee2015}.

\begin{acknowledgments}
This work was supported in part by the Foundational Questions Institute (FQXi) through Grant No. 2011-02814, the Canadian Institute for Advanced Research (CIFAR) through its Quantum Information Science Program and by the European Research Council (ERC) through project 257531 - EnSeNa. R.K. is supported via a Lise-Meitner-Fellowship of the Austrian Science Fund (FWF) (project M 1849). C.B. acknowledges support from the European Commission project RAQUEL (No. 323970); the Austrian Science Fund (FWF) through the Special Research Programme FoQuS, the Doctoral Programme CoQuS and Individual Project (No. 2462).
\end{acknowledgments}



%

\clearpage
\pagebreak
\onecolumngrid
\section{Supplemental Material}
\section{S1-Robustness of the Sorkin experiment}
\noindent In the following, we will outline the impact of typical error sources occurring in optical multi-path interferometers. For the sake of simplicity, we will restrict the analysis to potential second-order interference $\kappa_3$ arising across three paths $A$, $B$, $C$ with associated photon transmission rates into the target output mode $p_A$, $p_B$, $p_C$ and phases $\phi_A$, $\phi_B$, $\phi_C$. An extension to yet higher orders of interference is straightforwardly possible. There are four types of errors, which we are going to consider: 
\begin{itemize}
\item \indent Incoherence, e.g. due to phase fluctuation shorter than a single measurement or limited beam overlap on the recombining beam splitter
\item \indent Coherent phase fluctuations (phase fluctuations on a time scale longer than each measurement, but shorter than a measurement cycle) 
\item \indent Input power fluctuations
\item \indent Poissonian photon counting uncertainty (Shot noise).
\end{itemize} 
It should be reasonable to assume that all sources of errors are independent from one another, such that they can be treated separately and their individual influences on the Sorkin experiment can be added.\\\\
We first exclude the counting uncertainty and keep it for later. 
A general three-path interferometer has respective transmissions $T_{A,B,C}$ from the input along either of its three paths into the output mode:
\beq
p_A=T_Ap_1,\:p_B=T_Bp_2,\:p_C=T_Cp_3,
\nonumber
\eeq
where the input rates $p_{1,2,3}$ are allowed to be different random variables accounting for power fluctuations between the respective single-path measurement settings.\\
A two-path measurement of modes $A$ and $B$ takes place at another instance in time and is, thus, subjected to another input power $p_4$:
\beq
p_{AB}=T_Ap_4+T_Bp_4+2\sqrt{T_Ap_4T_Bp_4}\cos(\phi_B-\phi_A)=p_4(T_A+T_B+2\sqrt{T_AT_B}\cos\phi_1),
\label{Twopathterm}
\eeq
where the phase difference $\phi_1\equiv\phi_B-\phi_A$ can be treated as a single random variable. Similarly one can express the other two-path terms as: $p_{BC}=p_5(T_B+T_C+2\sqrt{T_BT_C}\cos\phi_2)$ and $p_{AC}=p_6(T_A+T_C+2\sqrt{T_AT_C}\cos\phi_3)$, with independent realizations of phase differences $\phi_2\equiv\phi_C-\phi_B$ and $\phi_3\equiv\phi_A-\phi_C$ (Of course $\left\langle\phi_1+\phi_2+\phi_3\right\rangle=0$ must hold in a stationary experiment). The two-mode interference term arises as the difference of the two-mode count rate and the two corresponding single-path rates:
\beq
I_{AB}\equiv p_{AB}-p_A-p_B=\left[p_4\left(T_A+T_B+2\sqrt{T_AT_B}\cos\phi_1\right)-T_Ap_1-T_Bp_2\right].
\label{TwomodeIF}
\eeq
The other two terms are defined accordingly.\\\\
The three-path measurement is influenced by yet another input power $p_7$ and three random realizations of the phases $\phi_A$, $\phi_B$, $\phi_C$, which we denote by $\phi_{4,5,6}$:
\beq 
p_{ABC}=p_7\left[T_A+T_B+T_C+2\left(\sqrt{T_AT_B}\cos\left(\phi_5-\phi_4\right)+\sqrt{T_BT_C}\cos\left(\phi_6-\phi_5\right)+\sqrt{T_AT_C}\cos\left(\phi_4-\phi_6\right)\right)\right].
\label{Threepathterm}
\eeq
In this case the three phases influence the measurement at the same time, so their differences can no longer be treated as being independent from one another. Of course, the background rate $p_0$ can also be affected by noise. However, as its magnitude is usually much smaller than the other rates, we neglect it in the following analysis. Therefore, the second-order interference term $\epsilon_3$ defined in Eq. (4) in the main text is modeled by seven random input powers $p_{1,\ldots,7}$ and six random phases $\phi_{1,\ldots,6}$.\\
In the experiment we operate at the point of fully constructive interference, that is $\left\langle\phi_j\right\rangle=0$ for all phases.
In the following we will evaluate the impact of the various error sources on the unnormalized term $\epsilon_3$ before considering the normalization by the sum of the first-order interference terms $\delta_3$.

\subsection{Phase fluctuations}
\noindent We first consider pure phase fluctuations, that is the absence of any changes in input power. Thus $p_i=p_{\mathrm{in}}=\mathrm{const.}$ for all $i=1,\ldots,7$ and one obtains for the second-order interference term:
\beq
\epsilon_3=2p_{\mathrm{in}}\left\{\sqrt{T_AT_B}\left[\cos\left(\phi_5-\phi_4\right)-\cos\phi_1\right]+\sqrt{T_BT_C}\left[\cos\left(\phi_6-\phi_5\right)-\cos\phi_2\right]+\sqrt{T_AT_C}\left[\cos\left(\phi_4-\phi_6\right)-\cos\phi_3\right]\right\}.
\label{ThirdOrderPhaseFluc}
\eeq
Whether phase fluctuations influence the experiment in a coherent or an incoherent way will depend on their time scale. We will evaluate both cases separately.

\subsubsection{Incoherence}
\noindent Incoherence can be modeled by rapid fluctuations of all phases around their mean value $\left\langle\phi\right\rangle=0$. This means that during a measurement the detector integrates over these rapid fluctuations reducing the result from the ideal value $\cos\phi=1$ to some averaged value $\left\langle\cos\phi\right\rangle=X$, with $0\leq X\leq 1$ measuring the degree of coherence and, thereby, the interference visibility. If the incoherence is stationary, it will reduce the interference contrast for all measurement settings by the same amount. As evident from Eq.~\ref{ThirdOrderPhaseFluc}, the cosines for each pair of paths enter with opposite signs, such that reduced interference contrasts cancel out and no net effect on $\epsilon_3$ remains. Thus, the Sorkin experiment is immune to such incoherence effects, as has been known for some time \cite{sollner12}.

\subsubsection{Coherent phase fluctuations}
\noindent Slower phase fluctuations preserve the phase within a single measurement, but can alter the phase inbetween measurements of a cycle. This can be modeled by assigning independent random values to the phases $\phi_{4,5,6}$ adhering to a Gaussian distribution with zero mean and a standard deviation of $\sigma_{\phi}$: $\phi_{4,5,6}\propto\mathcal{N}_{0,\sigma_{\phi}^2}$. The phase differences $\phi_1$, $\phi_2$, $\phi_3$ are differences of two such independent Gaussians. Hence, they are also normally distributed, but with twice the variance: $\phi_{1,2,3}\propto\mathcal{N}_{0,2\sigma_{\phi}^2}$. As in the scenario of incoherence, the phase fluctuations reduce the cosine terms. However, not every cosine is reduced in the same way, but by the six independent random phases. One can straightforwardly see from Eq.~(\ref{ThirdOrderPhaseFluc}) that there is no bias on the unnormalized second-order interference:
\beq
\left\langle \epsilon_3\right\rangle=0,
\nonumber
\eeq
as all terms are cosines of phase differences with identical spread, so they have identical expectation values which cancel each other out. Hence, also coherent phase fluctuations have no systematic influence on our experiment. They merely lead to random uncertainties which can be mitigated by averaging over multiple data sets.\\
\bfig
\centering
\includegraphics[width=160mm]{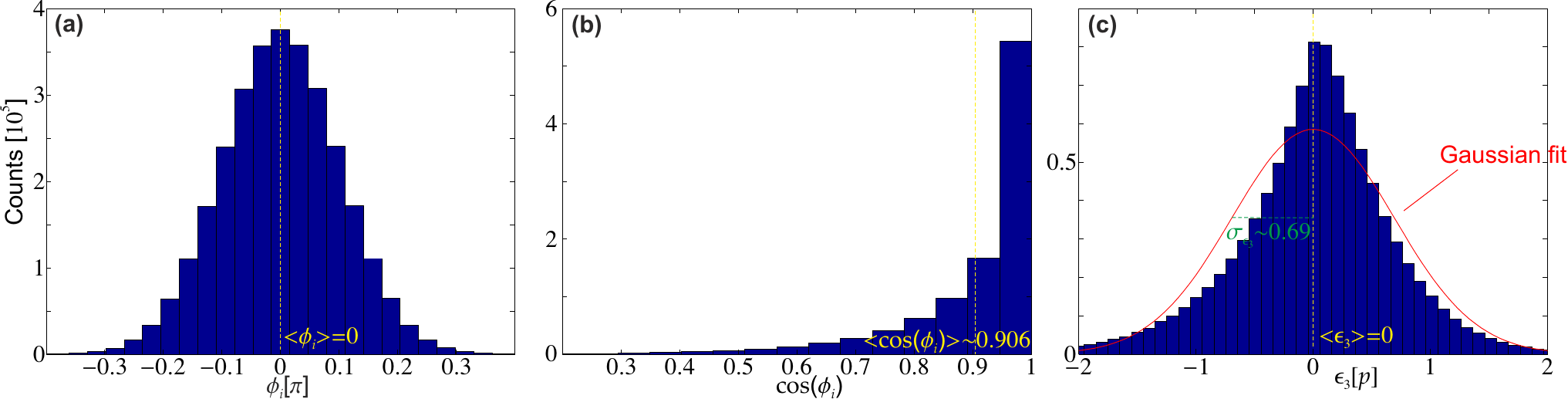}\caption{\label{SorkinPhaseError}Statistical data for a phase fluctuation $\sigma_{\phi}=0.1\pi$. \textbf{(a)} Distribution of the phase differences $\phi_{1,2,3}$. \textbf{(b)} Distribution of their cosine. \textbf{(c)} Distribution of $\epsilon_3$. The red curve is a normal distribution with zero mean value and standard deviation $\sigma_{\epsilon_3}$ for comparison. The ensemble size of the simulation was $10^6$.}
\efig
As the cosine is changing its value only in second order in $\phi$ at the maximum, one can expect a quadratic scaling of the uncertainty of $\epsilon_3$ with such phase fluctuations. We verify this intuitive expectation by numerical simulations. To this end, the random phases $\phi_{1,\ldots,6}$ are drawn from the Gaussian distributions defined above, with an ensemble size chosen large enough to ensure at least two significant digits for all moments in each case.
Fig.~\ref{SorkinPhaseError} shows simulated data for a phase uncertainty of $\sigma_{\phi}=\pi/10$ in a balanced interferometer ($T_A=T_B=T_C=T$; $p\equiv Tp_{\mathrm{in}}$ as the rate transmitted through a single path). As the cosine of a Gaussian variable centered around zero yields a highly asymmetric distribution (see subfigure \textbf{(b)}), the resulting distributions for the second-order interference are also asymmetric and deviate considerably from a normal distribution. Yet, the expectation value of $\epsilon_3$ remains exactly zero, as expected from our earlier considerations \textbf{(c)}.\\
In order to determine, the scaling of the random errors on $\epsilon_3$ with the level of phase noise, the procedure is repeated for a variety of phase uncertainty levels. The resulting standard deviations $\sigma_{\epsilon_{3,\mathrm{phase}}}$ are summarized in Table~\ref{PhaseErrorTable}:
\begin{table}[!ht]
\centering
\begin{tabular}{c||cccc}
$\delta\phi[\pi]$ & 0.1 & 0.03 & 0.01 & 0.003 \\
\hline
$\sigma_{\epsilon_{3,\mathrm{phase}}}[p]$ & 0.69 & 0.068 & $7.6\times 10^{-3}$ & $6.9\times 10^{-4}$\\
\end{tabular}
\caption{\label{PhaseErrorTable}Scaling of the uncertainty of $\epsilon_3$ with phase noise magnitude.}
\end{table}
For all cases, there is no bias. The scaling of the random uncertainty can be extracted from the table as being roughly
\beq
\sigma_{\epsilon_{3,\mathrm{phase}}}\propto 7.7p\sigma_{\phi}^2,
\nonumber
\eeq  
which is a quadratic scaling, as expected in the vicinity of fully constructive interference.

\subsection{Power fluctuations}
\noindent Now we consider the other extreme case of perfect phase stability but random fluctuations of the input power. Consequently, one can set all cosines in Eqs.~(\ref{Twopathterm}) and (\ref{Threepathterm}) to $1$ and treat the rates $p_{1,\ldots,7}$ as independent Gaussians with mean $p_{\mathrm{in}}$ and standard deviation $\sigma_p$. One obtains for the three-path interference term:
\begin{eqnarray}
\epsilon_3=T_A\left(p_7-p_4-p_6+p_1\right)+T_B\left(p_7-p_4-p_5+p_2\right)+T_C\left(p_7-p_5-p_6+p_3\right)+\ldots\nonumber\\
\ldots+2\left[\sqrt{T_AT_B}\left(p_7-p_4\right)+\sqrt{T_BT_C}\left(p_7-p_5\right)+\sqrt{T_AT_C}\left(p_7-p_6\right)\right].
\label{Eps3PowerFluc}
\end{eqnarray}
Again, no bias arises as the expectation values cancel each other to zero:
\beq
\left\langle\epsilon_3\right\rangle=0.
\nonumber
\eeq
Thus, the Sorkin experiment suffers no systematic error from power fluctuations, as long as all shutter settings receive the same expected input power. The influence of potential long-term drifts is eliminated by picking the shutter settings in random order for each measurement cycle.\\ 
As Eq.~(\ref{Eps3PowerFluc}) is linear in the random powers $p_{1,\ldots,7}$, the second-order interference follows also a Gaussian distribution with a standard deviation $\sigma_{\epsilon_{3,\mathrm{power}}}$ scaling linearly with $\sigma_p$. The coefficient depends on the specific values of the transmittivities. For a balanced interferometer, one obtains
\beq
\sigma_{\epsilon_{3,\mathrm{power}}}=2\sqrt{33}T\sigma_p=2\sqrt{33}p\frac{\sigma_p}{p_{\mathrm{in}}}.
\nonumber
\eeq

\subsection{Counting Error}
\noindent Finally, we take the photon counting noise into account. The number of incident photons in a given time interval follows Poissonian statistics, with equal mean photon number and variance. Hence, the count rate measured for shutter setting \lq A open\rq\ has a standard deviation of $\sqrt{p_A}$, and so on. Again, the second-order interference is unbiased by these uncertainties, as the mean values of the individual rates are unaffected. The counting errors for the different shutter settings are uncorrelated, so one can add their variances and obtain
\beq
\sigma_{\epsilon_{3,\mathrm{count}}}=\sqrt{p_{ABC}+p_{AB}+p_{BC}+p_{AC}+p_A+p_B+p_C}.
\nonumber
\eeq
In case of a balanced interferometer with absolute single path rate $p$ one gets in absence of other error sources
\beq
\sigma_{\epsilon_{3,\mathrm{count}}}=2\sqrt{6}\sqrt{p}.
\nonumber
\eeq
Other sources of noise in the detector can be modeled in the same way and also cause no bias on the result. 

\subsection{Combined uncertainty and normalization}
\noindent The above discussion shows that none of these error sources has a systematic effect on the unnormalized second-order interference. The three sources of random error can be expected to be independent from another, yielding a total uncertainty of
\beq
\sigma_{\epsilon_3}=\sqrt{\sigma_{\epsilon_{3,\mathrm{phase}}}^2+\sigma_{\epsilon_{3,\mathrm{power}}}^2+\sigma_{\epsilon_{3,\mathrm{count}}}^2}\approx 2\sqrt{3p}\sqrt{p\left(5\sigma_{\phi}^4+11\left(\sigma_p/p_{\mathrm{in}}\right)^2\right)+2},
\nonumber
\eeq
with the numbers in the last step applying to the case of a perfectly balanced interferometer. Evidently, for large count rates phase and power fluctuations, which enter in second and first order, respectively, tend to dominate over shot noise. Note that the phase noise does not lead to a Gaussian distribution of $\epsilon_3$, whereas the others do. Hence, histogram plots of $\epsilon_3$ can reveal which type of error source is dominant.
\\\\
It seems more useful to determine higher-order interference relative to first-order interference rather than in absolute terms.
Therefore, $\epsilon_3$ is normalized by the sum of the magnitudes of the first-order terms as defined in Eq.(\ref{TwomodeIF}):
\beq
\kappa_3\equiv\frac{\epsilon_3}{\delta_3}=\frac{\epsilon_3}{\left|I_{AB}\right|+\left|I_{BC}\right|+\left|I_{AC}\right|}.
\label{ThreeModeIFNorm}
\eeq
Clearly, the numerator and the denominator are dependent upon each other. Consequently, fluctuations in any of the random variables can lead to an undesired bias on $\kappa_3$.\\
In order to overcome this problem, we take the averages for $\epsilon$ and $\delta$ separately, as stated in Eq.~(5) in the main paper. Thereby, we obtain an unbiased estimate for $\epsilon$ (which is zero in the conventional theory) and normalize it by the averaged first-order interference. The latter is reduced by incoherence and phase fluctuations. The other error sources have no systematic influence on $\delta$.\\
The relative random uncertainty of the numerator $\epsilon_3$ is much larger than the one of the denominator $\delta_3$, due to $\left\langle\epsilon_3\right\rangle\approx 0$, but $\left\langle\delta_3\right\rangle >0$. Therefore, the standard deviation of $\kappa$
can be approximated by
\beq
\sigma_{\kappa_3}\approx\frac{\sigma_{\epsilon_3}}{\left\langle\delta_3\right\rangle}.
\label{kappaerror}
\eeq
In case of a perfectly balanced interferometer at fully constructive interference one can evaluate the denominator as $\left\langle\delta_3\right\rangle=6p$, thus:
\beq
\sigma_{\kappa_3}\approx\frac{1}{\sqrt{3}}\sqrt{\left(5\sigma_{\phi}^4+11\left(\sigma_p/p_{\mathrm{in}}\right)^2\right)+2/p}.
\nonumber
\eeq

\section{S2-Error analysis of the data}
\begin{figure}[!ht]
\centering
\includegraphics[width=\textwidth]{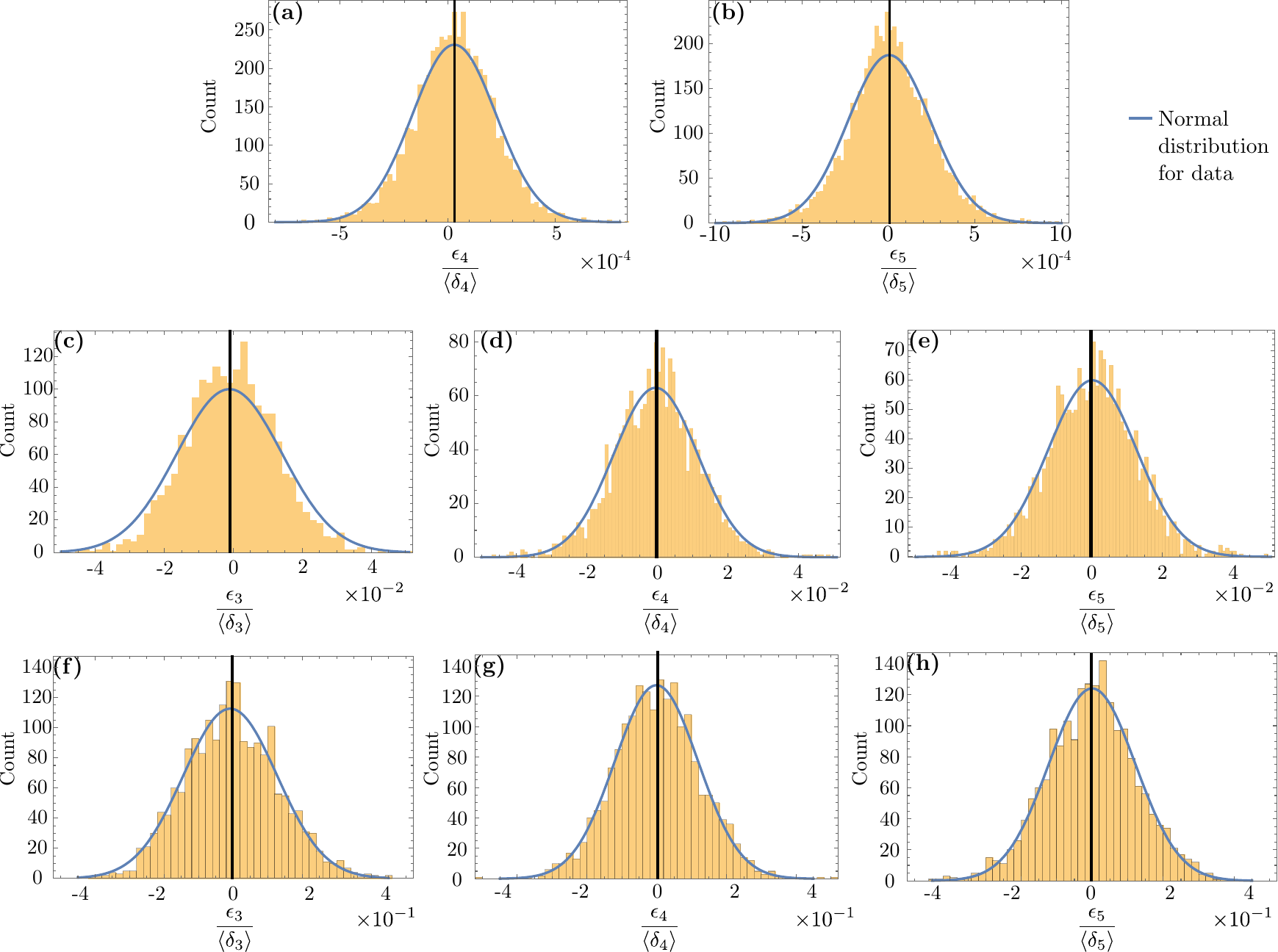}
\caption{Histograms for various orders of interference and measurement regimes. Top row: Data obtained in the classical regime with $M=5618$ measurements for paths $ABCD$ \textbf{(a)} and $ABCDE$ \textbf{(b)}. The corresponding three-path interference data is shown as Fig. 3 in the main paper. Middle row: Data obtained in the unheralded single-photon regime ($M=1912$) for paths $ABC$ \textbf{(c)} $ABCD$ \textbf{(d)} and $ABCDE$ \textbf{(e)}. Bottom row: Data obtained in the heralded single-photon regime ($M=1912$) for paths $ABC$ \textbf{(f)} $ABCD$ \textbf{(g)} and $ABCDE$ \textbf{(h)}. The blue lines are Gaussian fits of the expected distributions of the data and the black bars indicate the mean values with their widths as the standard error of mean.}
\label{kappahisto} 
\end{figure}
\noindent We conduct a series of $M$ measurements of $\epsilon$ and $\delta$, cycling through all shutter settings. The resulting distribution of $\epsilon$ for some exemplary path combinations, normalized by the average of $\delta$, is shown in Fig.~\ref{kappahisto} as well as in Fig. 3 in the main text.
As discussed in the previous section, the distribution of the data can provide information on the magnitude of random uncertainty as well as on the predominant source of error. In the classical and semi-classical regime the central peak of the distributions is slightly more pronounced than expected for an ideal Gaussian distribution. This is consistent with the expected shape of a phase-noise induced distribution (cf. Fig.~\ref{SorkinPhaseError}\textbf{(c)}). Therefore, the data suggests that both, Gaussian power and shot noise as well as non-Gaussian phase noise play a role in these regimes. For heralded single photons, on the other hand, the lower count rates render the data shot-noise dominated, causing the results to be normally distributed (see bottom row in Fig.~\ref{kappahisto}).\\\\
If the measurement results $\epsilon^{(i)}$ ($i=1,\ldots,M$) are mutually uncorrelated, one can use the standard error of mean to quantify the uncertainty of the average:
\beq
\Delta\epsilon\equiv\sigma_{\epsilon}/\sqrt{M},
\label{som}
\eeq
with $\sigma_{\epsilon}$ as the measured standard deviation of the data set. The same holds for the uncertainty of $\kappa$.
We verify the absence of correlations in the data by calculating their auto-correlation function: 
\beq
R_{\mathrm{ac}}\left(k\right)\equiv\frac{\sum_i\epsilon^{(i)}\epsilon^{(i+k)}}{\sum_i\left(\epsilon^{(i)}\right)^2}.
\nonumber
\eeq
\bfig
\centering
\includegraphics[width=160mm]{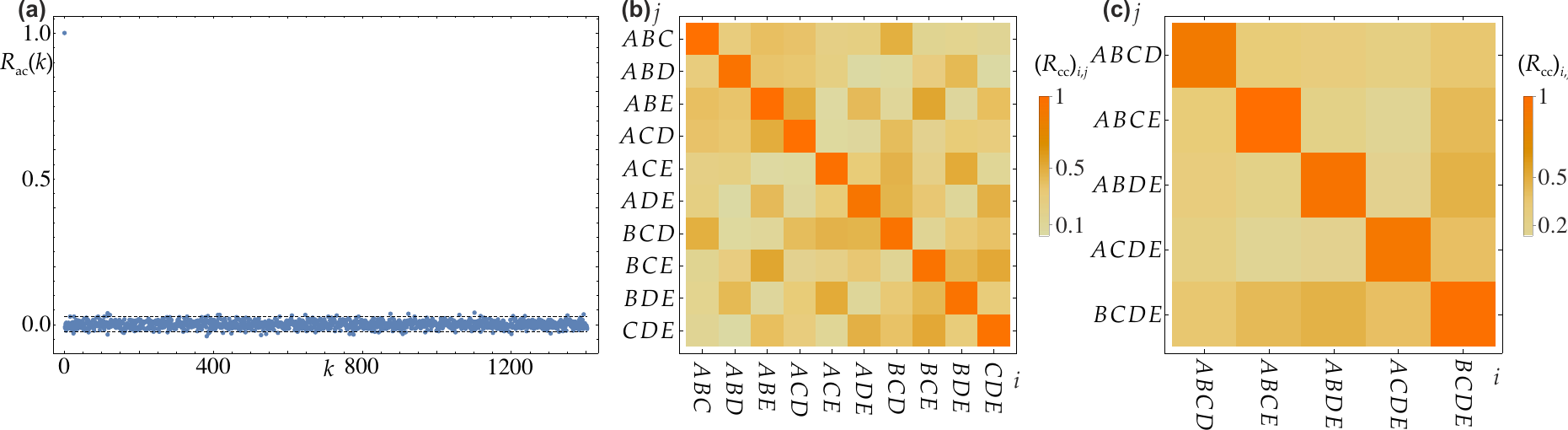}\caption{\label{Correlations}Auto- and crosscorrelation of the experimental data obtained with the laser source. \textbf{(a)} Autocorrelation of $\epsilon_3$ for path combination $ABC$. Only the first quarter of the correlation function is shown. For uncorrelated normally distributed data $95\%$ of the points should lie inbetween the dashed lines. \textbf{(b)} Crosscorrelation matrix between the second-order interference $\epsilon_3$ measured for the various path triplets. \textbf{(c)} Same for the third-order interference $\epsilon_4$.}
\efig
An exemplary auto-correlation for $\epsilon_3$ measured with the laser source on paths $A,B,C$ is shown in Fig.~\ref{Correlations}\textbf{(a)}. Evidently, no significant correlation among the data points can be detected, not even for subsequent ones. The same observation is made for all other measurement settings and light sources. Therefore, it is fully justified to use the standard error of mean, as defined in Eq.~(\ref{som}), as a measure for the uncertainty of the higher-order interference.\\\\
One has to be more careful, however, when it comes to averaging across the various path combinations in the multi-path interferometer. For example, phase fluctuations in a single path will influence the interference among several path combinations simultaneously. Therefore, cross-correlations between the measured data for different path combinations ($i=ABC,\ldots,CDE$ for three paths) can be expected. Indeed, one finds non-vanishing cross-correlations $\left(R_{\mathrm{cc}}\right)_{i,j}$ for $\epsilon_3$ and $\epsilon_4$, as shown in Fig.~\ref{Correlations}\textbf{(b)} and \textbf{(c)}, respectively. With this at hand, one can obtain the standard error of mean of the final result for $\kappa_3$ via error propagation:
\beq
\Delta\kappa_3=\frac{1}{10}\sqrt{\sum_{i,j=ABC}^{CDE}\left(R_{\mathrm{cc}}\right)_{i,j}\Delta\kappa_{3,i}\Delta\kappa_{3,j}},
\nonumber
\eeq
with $\Delta\kappa_{3,i}$ denoting the standard error of mean of $\kappa_3$ on path combination $i$. Analogously, one gets for $\kappa_4$:
\beq
\Delta\kappa_4=\frac{1}{5}\sqrt{\sum_{i,j=ABCD}^{BCDE}\left(R_{\mathrm{cc}}\right)_{i,j}\Delta\kappa_{4,i}\Delta\kappa_{4,j}}.
\nonumber
\eeq
These are the experimental uncertainties presented in Table I in the main text.

\section{S3-Tomography data and predicted count rates}

For completeness we show additional data/figures resulting from our measurement:

The measured count rates of the different path combinations for the measurement with the laser can be seen in Fig.~\ref{fig:intensitysp}. After filtering the 5618 measurement sets recorded in a measurement time of \SI{68}{\hour} the standard deviation was measured to be 0.3\%. A bar chart with the average powers over the whole measurement duration can be seen in Fig.~2 of the main text.

\begin{figure}[!ht]
\centering
\includegraphics[width=\textwidth]{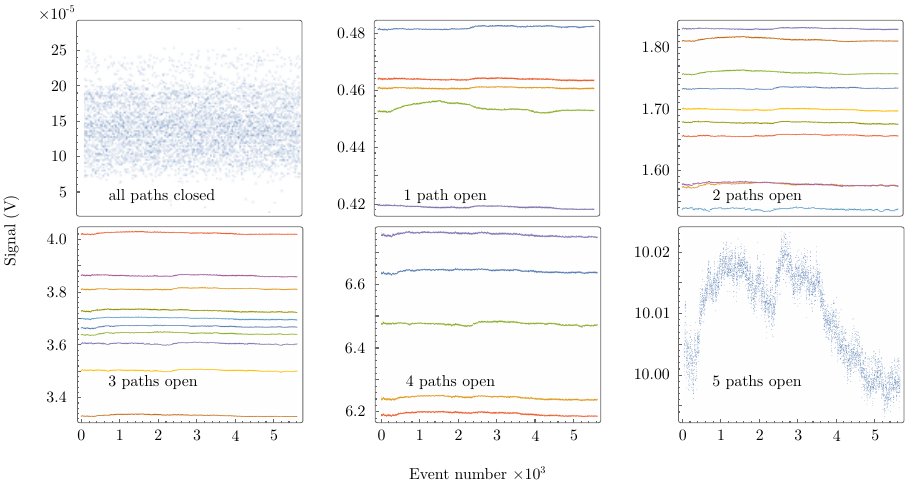}
\caption{Measured intensities for the different path-combinations in the classical regime. Each combination is indicated by a separate color within each panel.}
\label{fig:intensitysp}
\end{figure}

The measured count rates of the different path combinations for the measurement with the single photon source for the semi-classical measurement can be seen in Fig.~\ref{fig:intensitysp}. After filtering the 1912 measurement sets recorded in a measurement time of \SI{88}{\hour} the standard deviation was measured to be 3.6\%. This higher standard deviation compared to the classical light source results mainly from shot noise and due to the fact that the power of the blue pump laser was not stabilized.

\begin{figure}[!ht]
\centering
\includegraphics[width=\textwidth]{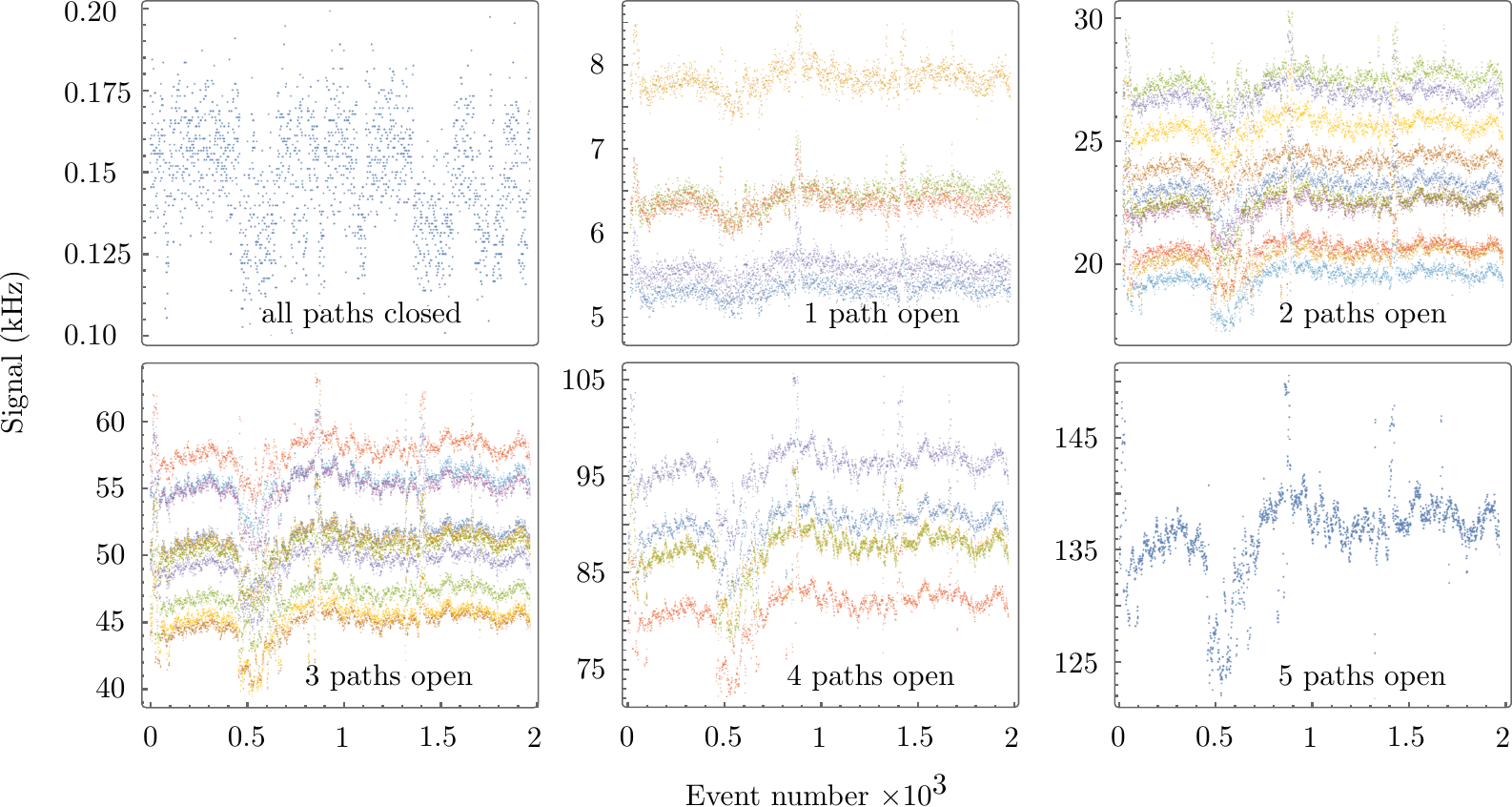}
\caption{Measured intensities for the different path-combinations in the semi classical regime. Each combination is indicated by a separate color within each panel.}
\label{fig:intensitysp}
\end{figure}

\begin{figure}[!ht]
\centering
\includegraphics[width=\linewidth]{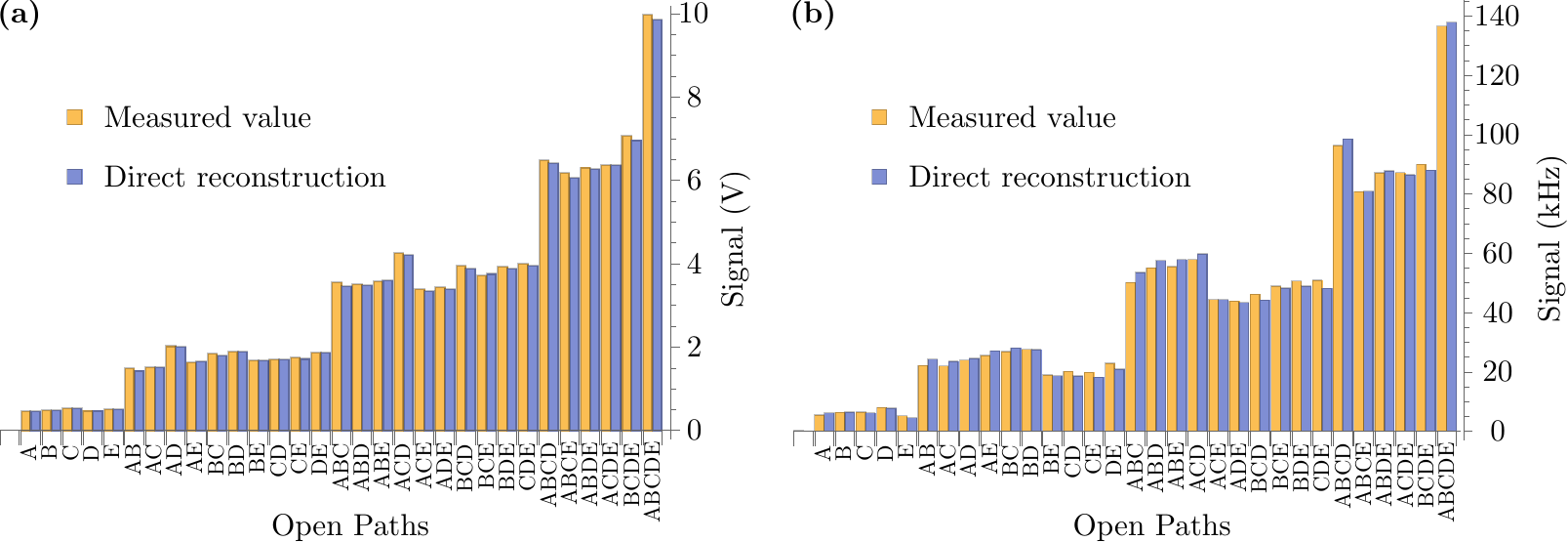}
\caption{Comparison between measured data and data simulated from tomography for \textbf{(a)} the classical and \textbf{(b)} semi-classical measurement.}
\label{fig:comparison}
\end{figure}

\noindent Fig. \ref{fig:comparison} shows the comparison between the measured values (shown in Fig. 2 in the main text for the laser) and the expected values arising from the reconstructed density matrix for all different path combinations.

\section{S4-Nonlinearity measurement and associated uncertainty}
\noindent Detector nonlinearities are the predominant cause of systematic deviations in our implementation of the Sorkin experiment. We deduce these nonlinearities from independent beam combination experiments \cite{Coslovi1980,kauten14}. Their expected impact on the higher-order interference term $\kappa_{\mathrm{th}}$ is then calculated by applying the nonlinear detector response function to the various multi-path probabilities calculated from the density matrix of the interferometer. In the following, it will be discussed how and with which uncertainty the nonlinearities are reconstructed and how this translates into the theoretical prediction $\kappa_{\mathrm{th}}$ and the error bars of this prediction.
\subsection{Nonlinearity of the photodiode and the voltmeter}
\subsubsection{Polynomial expansion of the transfer function}
\noindent In a measurement instrument for classical light the output voltages $V$ are related to the impinging optical powers $p$ by some transfer function $p=f(V)$. A variety of effects can contribute to nonlinearities in a photodiode and the attached voltmeter, such that it is difficult to develop a specific model. However, as the nonlinearities are quite weak compared to the linear part of the transfer function (on the $\unit[10]{ppm}$-level), one can reasonably expand $f(V)$ into a polynomial of degree $n$ in the region of interest \cite{kauten14}:
\beq
f(V)=\sum_{j=0}^n a_jV^j+r_n(V,\mathbf{a}),
\nonumber
\eeq
with coefficients $\mathbf{a}\equiv\left(a_0,\ldots,a_n\right)^T$ and the residual $r_n(V,\mathbf{a})$ quantifying the fidelity of this approximation for a given measured voltage $V$ and polynomial degree $n$.\\\\
The goal of the beam-combination method is to find a suitable polynomial for $f(V)$. One overlaps two incoherent beams, which can be shuttered and power-controlled individually, on a detector (see, e.g., Fig. 1 in \cite{kauten14} for an illustration). For a series of $k=1,\ldots,M$ settings of optical powers, the voltages $V_{1,k}$, $V_{2,k}$ for either of the paths being open, $V_{3,k}$ for both path being open and $V_{0,k}$ for both paths being closed are recorded. The corresponding unknown optical powers are $p_{0,k},\ldots,p_{3,k}$. For incoherent beams, the optical powers should be additive, i.e.,
\beq
\forall k:\:p_{3,k}+p_{0,k}=p_{1,k}+p_{2,k}.
\label{poweradditivity}
\eeq
Condition~(\ref{poweradditivity}) leads to the following constraints on the coefficients:
\beq
\forall k:\sum_{j=1}^na_j\left(V_{0,k}^j+V_{3,k}^j-V_{1,k}^j-V_{2,k}^j\right)=r_n(\mathbf{V_k},\mathbf{a}),
\label{constraint}
\eeq
with $r_n(\mathbf{V}_k,\mathbf{a})=r_n(V_{1,k},\mathbf{a})+r_n(V_{2,k},\mathbf{a})-r_n(V_{3,k},\mathbf{a})-r_n(V_{0,k},\mathbf{a})$ measuring the combined residual for data set $k$. Note that the coefficient $a_0$ drops out of the sum due to $V_{0,k}^0=\ldots=V_{3,k}^0=1$. This is to be expected, as constant off-sets should have no influence in a balanced zero-sum experiment as given by Eq.~(\ref{poweradditivity}).\\\\ 
The aim is to determine the nonlinearity of the detector. The exact value of the linear slope $a_1$ is of no interest in this respect. Thus, it can be set to $a_1=1$ for simplicity. 
It is convenient to introduce the quantity
\beq
S_{j,k}\equiv V_{0,k}^j+V_{3,k}^j-V_{1,k}^j-V_{2,k}^j.
\nonumber
\eeq
With this at hand, one can reformulate the constraints~(\ref{constraint}) into the following least-square optimization problem:
\beq
R_n(\mathbf{\bar{a}})\equiv\sum_{k=1}^M{\frac{\left|S_{1,k}+\sum_{j=2}^na_jS_{j,k}\right|^2}{\sigma_k^2}};\;R_n(\mathbf{\bar{a}})\rightarrow\mathrm{Min},
\label{LeastSquares}
\eeq
which is solved by finding a configuration of nonlinearity parameters $\mathbf{\bar{a}}\equiv\left(a_2,\ldots,a_n\right)^T$ that minimize $R_n(\mathbf{\bar{a}})$. Here, the contribution of each dataset $k$ is inversely weighted with its variance $\sigma_k^2$. These variances are a result of random experimental uncertainties of the measured voltages and may vary across the measuring range of the instrument. Taking them into account bears the advantage that more certain parts of the data contribute more to the optimization than uncertain ones.\\ 
If these data uncertainties are known, one can straightforwardly solve the least-square problem \cite{Richter:ErrorLeastSquares}: 
\beq
\mathbf{\bar{a}}=\mathbf{C}\mathbf{A}^T\mathbf{b},
\nonumber
\eeq
with the covariance matrix
\beq
\mathbf{C}\equiv\left(\mathbf{A}^T\mathbf{A}\right)^{-1},
\label{Covmat}
\eeq
the weighted matrix elements $A_{k,j-1}\equiv S_{j,k}/\sigma_k$ (for $k=1,\ldots,M$ and $j=2,\ldots,n$) and vector components $b_k\equiv-S_{1,k}/\sigma_k$. Then, the optimal transfer function reads
\beq
f_{\mathrm{opt}}(V)=V+\sum_{j=2}^na_jV^j,
\label{ResultTransferFunction}
\eeq
where the irrelevant off-set term $a_0$ has been set to zero for convenience.

\subsubsection{Uncertainty of the nonlinearity}
\noindent In the following, it will be explained how the variances $\sigma_k^2$ can be extracted from the data and how they influence the uncertainty of the optimization. The experimental data of \cite{kauten14} is used for this procedure and the subsequent prediction of $\kappa_{\mathrm{th}}$.\\
The optical power in the beam combination experiment is increased linearly between subsequent measurement points (see Fig.~\ref{NLData}\textbf{(a)}). The uncertainty of each voltage $V_{m,k}$ is estimated by calculating a floating standard deviation $\sigma_{V_{m,k}}$ of the surrounding data points ($r=101$ points were taken in each case) with respect to a perfectly linear slope (Fig.~\ref{NLData}\textbf{(b)}).
The uncertainty $\sigma_k$ can then be taken as the standard deviation of each summand in Eq.~(\ref{LeastSquares}), i.e., as the standard deviation of the function $Y(\mathbf{V_k},\mathbf{\bar{a}})\equiv S_{1,k}+\sum_{j=2}^n a_jS_{j,k}$. If the 4 voltages in each dataset $\mathbf{V_k}$ are independent from one another and if the uncertainties of the voltages are small compared to their mean values, one can express $\sigma_k^2$ as \cite{Clutton-Brock:EstimatingFunctionsDoubleError}:
\beq
\sigma_k^2=\sum_{m=0}^3\left(\frac{\partial Y(\mathbf{V_k},\mathbf{\bar{a}})}{\partial V_{m,k}}\right)^2 \sigma_{V_{m,k}}^2=\left(1+\sum_{j=2}^nja_jV_{m,k}^{j-1}\right)^2\sigma_{V_{m,k}}^2.
\nonumber
\eeq
In general, this depends on the measured voltages as well as on the nonlinearity coefficients. However, for a weak nonlinearity $a_jV^j\ll V$ holds for all $j>1$. Hence, $a_jV^{j-1}\ll 1$ and $\sigma_k$ is dominated by uncertainties in the linear part of the optimization problem:
\beq
\sigma_k^2\approx\sum_{m=0}^3\sigma_{V_{m,k}}^2.
\nonumber
\eeq
\bfig
\centering
\includegraphics[width=170mm]{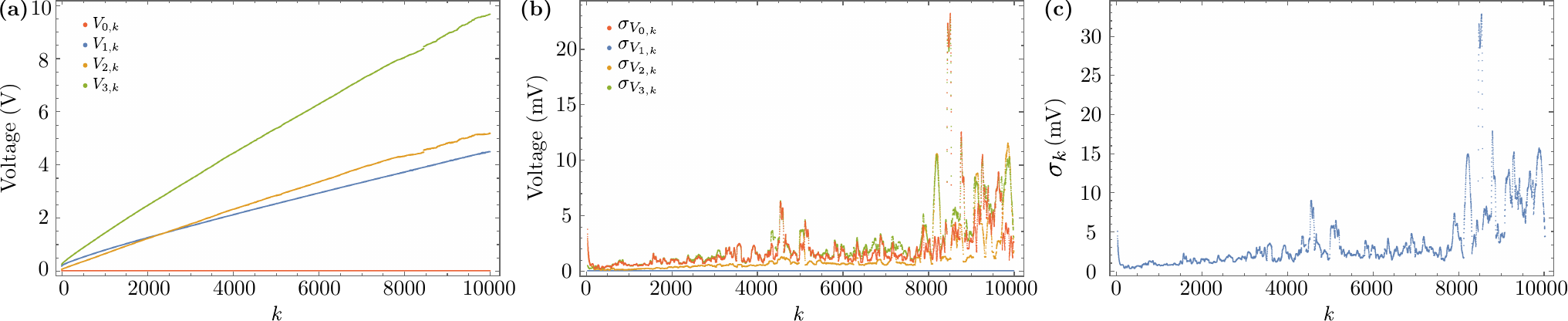}
\caption{\label{NLData} \textbf{(a)} Measured voltages in the beam combination experiment. \textbf{(b)} Uncertainties of these voltages, obtained from a floating standard deviation. \textbf{(c)} Total uncertainty of the dataset, which is used as weight in the least-square optimization.}
\efig
The resulting uncertainties are shown in Fig.~\ref{NLData}\textbf{(c)} and are used as weights in the least-square optimization~(\ref{LeastSquares}).\\\\ 
The variances and covariances of the resulting coefficients $\mathbf{\bar{a}}$ can then be readily obtained from the covariance matrix~(\ref{Covmat}) \cite{Richter:ErrorLeastSquares}:
\beq
\sigma_{a_j}^2=C_{j-1,j-1};\;\mathrm{Cov}(a_j,a_k)=C_{j-1,k-1}.
\nonumber
\eeq
The uncertainty of the reconstructed transfer function in Eq.~(\ref{ResultTransferFunction}) at a given voltage $V$ can be calculated from the covariance matrix via error propagation
\beq
\sigma_f^2(V)=\sum_{p,q=2}^n \frac{\partial f_{\mathrm{opt}}}{\partial a_p}\frac{\partial f_{\mathrm{opt}}}{\partial a_q}C_{p-1,q-1}=\sum_{p,q=2}^n V^{p+q}C_{p-1,q-1}.
\label{functionuncertainty}
\eeq
Note that this purely data-based approach makes no assumption on the precision of the instruments or the magnitude of random fluctuations in the experimental conditions.

\subsubsection{Order of the polynomial and resulting transfer function}
\noindent The optimization procedure Eq.~(\ref{LeastSquares}) is solved for various degrees $n$ of the polynomial. We select the lowest possible $n$ beyond which the quality of the optimization does not significantly increase any further. To be specific, we calculate a test quantity $X(n)$, defined as the residual $R_n(\mathbf{\bar{a}})$, normalized by the remaining number of degrees of freedom in the system $X(n)\equiv R_n(\mathbf{\bar{a}})/(M-(r-1)-n+1)$,
with $M-(r-1)$ being the number of data points remaining after calculation of the local standard deviation.\\
\bfig
\centering
\includegraphics[width=160mm]{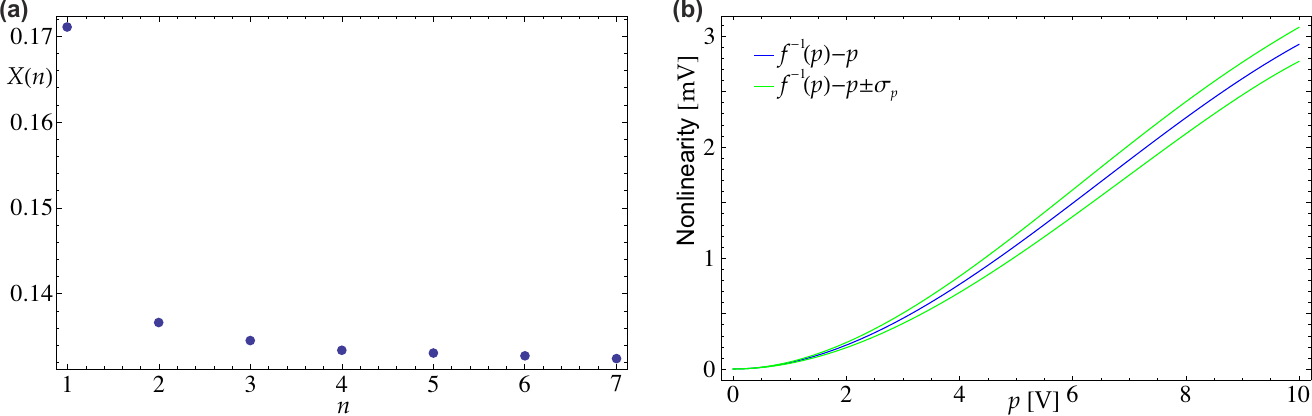}
\caption{\label{Orderscan} \textbf{(a)} Normalized residual $X(n)$ of the least square optimization in dependence of the chosen order of the polynomial $n$. \textbf{(b)} Resulting nonlinearity and associated uncertainty for $n=3$.}
\efig
This quantity has been calculated for increasing $n$ and the result is plotted in Fig.~\ref{Orderscan}\textbf{(a)}. As one would expect, a linear function $n=1$ fares worse than nonlinear ones, which suggests that indeed significant nonlinearity is present in the system. There is some improvement in the quality of the fit from $n=2$ to $n=3$ ($X(3)/X(2)\approx 0.98$). However, beyond $n=3$, the optimization improves barely any further: $X(4)/X(3)\approx 0.992$ and $X(5)/X(4)\approx 0.998$. This suggests that $n=3$ is a reasonable choice for the order of the polynomial without overinterpreting the data. This value is chosen for the estimation of the nonlinearity in this work.\\
The resulting parameters, which optimize~(\ref{LeastSquares}) for the beam combination measurements of Ref.~\cite{kauten14} are:
\beq
\mathbf{\bar{a}}=\left(-59.8,3.1\right)^T\times 10^{-6},
\eeq
with the covariance matrix
\beq
C=\left(\begin{array}{c c}
496 & -45\\
-45 & 4.2
\end{array}\right)\times 10^{-13}.
\eeq 

\subsubsection{Implications on the Sorkin experiment}
\noindent Now it will be discussed how the systematic influence of nonlinearity-affected detection of the interferometer's output powers on higher-order interferences can be quantified. To this end, the state of the interferometer, i.e., all amplitudes and phases of the 5 paths, is reconstructed via state tomography. From this the expected optical powers of the Sorkin terms $\left(p_A,p_{AB},\ldots\right)$ are calculated (see Fig.~\ref{fig:comparison}). Due to the nonlinear dependence between these powers and their measured voltages $\left(V_A=f_{\mathrm{opt}}^{-1}(p_A),V_{AB}=f_{\mathrm{opt}}^{-1}(p_{AB}),\ldots\right)$ an apparent non-zero higher order interference $\kappa_{3,\mathrm{th}}$ arises, even if no physical higher-order interference is present. Its value will be predicted by calculating the nonlinearity affected voltages via the inverse transfer function.\\
Clearly, any uncertainty in $f_{\mathrm{opt}}^{-1}(p)$ will translate directly into an error bar of the predicted $\kappa_{\mathrm{th}}$. The inverse function can be analytically calculated from $f_{\mathrm{opt}}(V)$ for low-order polynomials. However, the measurement device is mostly linear, such that $p=f(V)=V+f_{\mathrm{NL}}(V)$, with $\left|f_{\mathrm{NL}}(V)\right|\ll V$. With this one can approximate
\beq
f^{-1}(p)\approx p-f_{\mathrm{NL}}(p)=2p-f(p).
\label{Inversefunctionapproximation}
\eeq
Then, clearly, the uncertainty of the inverse function is approximately equal to the one of the forward transfer function.
\beq
\sigma_{f^{-1}}(p)\approx\sigma_f(p),
\nonumber
\eeq
which can be calculated via~(\ref{functionuncertainty}). The inverse transfer function and its uncertainty are shown in Fig.~\ref{Orderscan}\textbf{(b)}.\\\\
Note that the tomography data itself will also be affected by the nonlinearity, so one could apply the function $f_{\mathrm{opt}}(V)$ to reconstruct the \lq true \rq optical powers before the density matrix is reconstructed. However, the higher-order interference must always be zero for any physical density matrix, regardless of its precise structure (if Born's rule holds). Therefore, a forward nonlinearity correction of the tomography data, leading only to minuscule changes in the density matrix, has no significant impact on $\kappa_{\mathrm{th}}$ and can, hence, be omitted.\\\\
It is of course difficult to tell in which way the realizations of $f^{-1}(p)$ for the various values of $p$ are correlated with each other. One may consider two extreme cases: The case of \lq maximum correlation\rq, where all values of $f_{\mathrm{opt}}^{-1}(p)$ are shifted in the same direction by one standard deviation and the completely uncorrelated case, where each value $f_{\mathrm{opt}}^{-1}(p)$ is replaced by an independent Gaussian random variable with mean $f_{\mathrm{opt}}^{-1}(p)$ and standard deviation $\sigma_f(p)$.
In the maximally correlated scenario, one finds generally much smaller error intervals than in the uncorrelated case. Therefore, we resort here to the latter scenario to provide the more conservative error estimation.\\
For such a fully independent random variation of the nonlinearity for each power term, one can calculate the uncertainty of the prediction of $\kappa_{\mathrm{th}}$ by propagation of Gaussian errors. 
With this procedure one obtains for the dataset of the main paper and the corresponding tomography data, the apparent higher order interferences shown as the red symbols in Fig.~\ref{kappapaths}. Similar values and uncertainties are expected for all subsets of the five interferometer-paths. In case of second- and third-order interference, one can average over these subsets to obtain, along with the prediction for the fourth-order interference on the full set of paths, the final estimates used in the main text of the paper:
\bfig
\centering
\includegraphics[width=170mm]{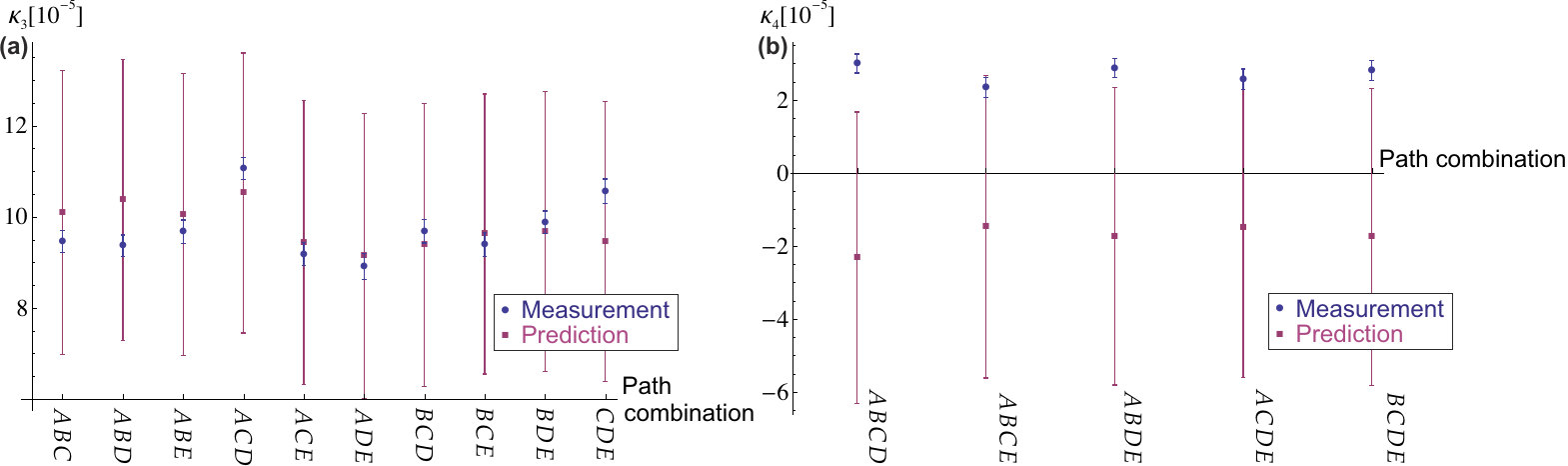}
\caption{\label{kappapaths}Experimentally measured higher-order interferences (blue data points and error bars) in comparison to the predicted values $\kappa_{\mathrm{th}}$ arising from detector nonlinearities (red). All possible path combinations are shown. Errors bars indicate one standard deviation. \textbf{(a)} Second-order interference $\kappa_3$. \textbf{(b)} Third-order interference $\kappa_4$.}
\efig
\beq
\begin{split}
\kappa_{3,\mathrm{th}}^{\mathrm{C}}=\left(9.7\pm 3.1\right)\cdot 10^{-5}\\
\kappa_{4,\mathrm{th}}^{\mathrm{C}}=\left(-1.6\pm 4.1\right)\cdot 10^{-5}\\
\kappa_{5,\mathrm{th}}^{\mathrm{C}}=\left(-3.9\pm 5.1\right)\cdot 10^{-5}.
\nonumber
\end{split}
\eeq
The uncertainties of the measured value $\langle\kappa\rangle$ and the predicted value $\kappa_{\mathrm{th}}$ arise from independent experiments. Therefore, it is natural to assume that there is no correlation between them, which implies that the uncertainty of the corrected higher-order interference $\tilde{\kappa}=\langle\kappa\rangle-\kappa_{\mathrm{th}}$ is obtained by adding the individual variances:
\beq
\Delta\tilde{\kappa}=\sqrt{\Delta\kappa^2+\Delta\kappa_{\mathrm{th}}^2}.
\nonumber
\eeq
These are the error bars shown in Table II and Fig. 5 in the main text.

\subsection{Nonlinearity of a single photon detector}
\noindent The reverse-biased avalanche diodes used as single photon detectors have an intrinsic deadtime $\tau$. The resulting saturation of the detector for increasing count rates is the dominant nonlinear mechanism. The transfer function $p=f(V)$ ($p$ and $V$ now being count rates) and its inverse can be expressed by a simple saturation model \cite{kauten14}:
\beq
f(V)=\frac{V}{1-\tau V};\;f^{-1}(p)=\frac{1}{1+\tau p}.
\label{Saturation}
\eeq
Just as for classical light detectors, the unknown parameter $\tau$ can be determined via beam combination experiments. The additivity of the optical powers~(\ref{poweradditivity}) translates to the following condition:
\begin{eqnarray}
\forall k:\:F(\tau,\mathbf{V_k})\equiv && f^{-1}\left[f\left(V_{0,k}\right)+f\left(V_{3,k}\right)\right]-f^{-1}\left[f\left(V_{1,k})+f(V_{2,k}\right)\right]\nonumber\\
=&&\frac{V_{0,k}+V_{3,k}-2\tau V_{0,k}V_{3,k}}{1-\tau^2 V_{0,k}V_{3,k}}-\frac{V_{1,k}+V_{2,k}-2\tau V_{1,k}V_{2,k}}{1-\tau^2 V_{1,k}V_{2,k}}=0.
\nonumber
\end{eqnarray}
The optimal solution for the deadtime is found via the nonlinear optimization problem:
\beq
R(\tau)\equiv\sum_{k=1}^M\frac{\left[F(\tau,\mathbf{V_k})\right]^2}{\sigma_k^2};\;R(\tau)\rightarrow\mathrm{Min}.
\eeq
As before, $\sigma_k^2$ is the variance of each function $F(\tau,\mathbf{V_k})$ and is used to weight the optimization. If one is sufficiently far away from the saturation point, the nonlinearity is weak and the variance is dominated by the uncertainty in the linear part of $F(\tau,\mathbf{V_k})$. Then, $\sigma_k^2\approx\sum_{m=0}^3\sigma_{V_{m,k}}^2$, just as for the classical detector. As before, the individual uncertainties $\sigma_{V_{m,k}}$ can be calculated from the data via a floating standard deviation.\\\\
Using this technique, we obtain a deadtime $\tau=\SI{33.9}{\nano\second}$ with an uncertainty $\sigma_{\tau}=\SI{0.3}{\nano\second}$ for the detectors used in our experiment. 
\bfig
\centering
\includegraphics[width=170mm]{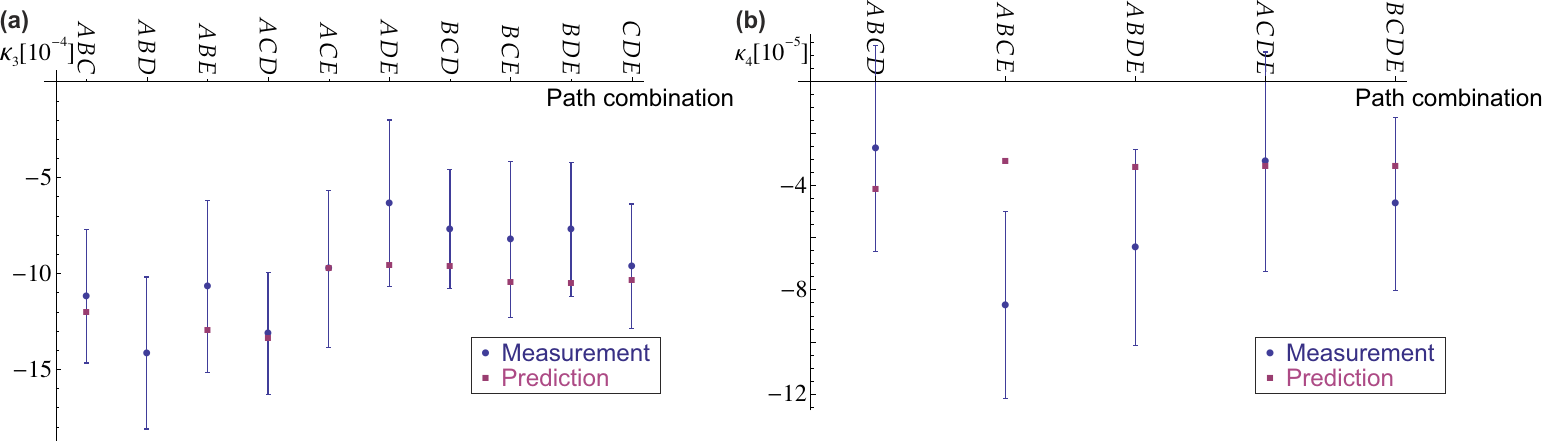}
\caption{\label{kappapathssp}Same as Fig.~\ref{kappapaths}, but for the single photon data. Error bars of the prediction are invisible on this scale.}
\efig
One can then use the saturation model~(\ref{Saturation}) to calculate a prediction of higher-order interference from the tomography data, just as for the classical light measurements. In this case, however, the uncertainty in $\tau$ must influence all count rates in the same direction. Therefore, the uncertainty in the prediction can be estimated by performing it also for deadtimes $\tau\pm\sigma_{\tau}$. These predictions are shown together with the experimental data in Fig.~\ref{kappapathssp} for the individual path combinations. Compared to the classical case, they have negligible error bars, as a very specific nonlinearity model with a single, precisely determined parameter is used. The averages over all path combinations yield:
\beq
\begin{split}
\kappa_{3,\mathrm{th}}^{\mathrm{SC}}=\left(-11.18\pm 0.10\right)\times 10^{-4}\\
\kappa_{4,\mathrm{th}}^{\mathrm{SC}}=\left(-3.48\pm 0.03\right)\times 10^{-4}\\
\kappa_{5,\mathrm{th}}^{\mathrm{SC}}=\left(2.85\pm 0.05\right)\times 10^{-6}.
\nonumber
\end{split}
\eeq

\subsection{Nonlinearity for the heralded single photons}
\noindent In case of heralded single photon detection one has to take into account that both, the heralding detector ($\mathrm{h}$) as well as the detector fed by the interferometer ($\mathrm{i}$) are inactive for the dead time $\tau$ whenever they detect a photon. Some of these cases are coincidences ($\mathrm{c}$), which blind both detectors. One can derive a saturation model, with the \lq true\rq\ coincidence rate $p_{\mathrm{c}}$ depending on the measured rates $V_{\mathrm{i,h,c}}$:
\beq
p_{\mathrm{c}}=f(V_{\mathrm{i}},V_{\mathrm{h}},V_{\mathrm{c}})=\frac{V_{\mathrm{c}}}{1-\tau \left(V_{\mathrm{i}}+V_{\mathrm{h}}-V_{\mathrm{c}}\right)}.
\nonumber
\eeq
This model is applied to correct the coincidence data obtained in the heralded single photon regime, which is then used to directly calculate the corrected higher order interference $\tilde{\kappa}^{\mathrm{HSP}}$ presented in Eq.~(11) in the main text. As for the unheralded single photons, the error interval of the correction is estimated by performing it with deadtimes $\tau\pm\sigma_{\tau}$. Again, this correction uncertainty is assumed to be independent from the experimental errors in the Sorkin measurement, such that their variances can be added.

\end{document}